\documentstyle[emulateapj,psfig]{article}

\newcommand{\be}{\begin{equation}}
\newcommand{\ee}{\end{equation}}
\newcommand{\ba}{\begin{eqnarray}}
\newcommand{\ea}{\end{eqnarray}}
\newcommand{\siml}{\lower4pt \hbox{$\buildrel < \over \sim$}}
\newcommand{\simg}{\lower4pt \hbox{$\buildrel > \over \sim$}}
\def\Mesz{M\'esz\'aros~}

\slugcomment{Accepted for publication in ApJ Letters}

\begin{document}

\title{Quasi-universal Gaussian jets: a unified picture for
gamma-ray bursts and X-ray flashes}

\author{Bing Zhang\altaffilmark{1}, Xinyu Dai\altaffilmark{1},
Nicole M. Lloyd-Ronning\altaffilmark{2} and Peter
\Mesz\altaffilmark{1,3,4} }

\altaffiltext{1}{Department of Astronomy \& Astrophysics, Pennsylvania
State University, University Park, PA 16802.}
\altaffiltext{2}{Los Alamos National Laboratory, MS B244, Los Alamos,
NM 87544.}
\altaffiltext{3}{Department of Physics, Pennsylvania State University,
University Park, PA 16802.}
\altaffiltext{4}{On leave at The Institute for Advanced Study,
Princeton, NJ 08540.}

\begin{abstract}
An observed correlation $E_p \propto (E_{iso})^{1/2}$ extending from
Gamma-ray Bursts (GRBs) to X-ray flashes (XRFs) poses problems both
for a power-law universal jet model where the energy per solid angle
decreases as the inverse square of the angle respect to the jet axis,
and for a conical jet model with a uniform energy density within the jet
beam and a sharp energy cut-off at the jet edge. Here we show that the
current GRB-XRF prompt emission/afterglow data can be understood in terms
of a picture in which the GRB-XRF jets are quasi-universal and structured,
with a Gaussian-like or similar structure, i.e., one where the jet has
a characteristic angle, with a mild variation of energy inside and a
rapid (e.g. exponential) decrease of energy outside of it.
A Monte Carlo simulation shows that the current data is compatible
with such a quasi-universal Gaussian jet with a typical opening angle of
$5.7^{+3.4}_{-2.1}$ degrees, and with a standard jet energy of about
$\log (E_j/\mbox{1 erg})= 51.1\pm 0.3$. According to this model, the
true-to-observed number ratio of the whole GRB-XRF population is about
14 with the current instrumental sensitivity.
\end{abstract}

\keywords{Gamma ray bursts - X-rays - gamma rays }

\section{Introduction}

The achromatic steepening in the afterglow lightcurves of some GRBs
as well as energy budget arguments suggest that GRBs are produced
by collimated jets (Rhoads 1999; Kulkarni et al. 1999; Harrison
et al. 2001). This suggestion receives indirect support from
the intriguing fact that a geometry-corrected jet energy appears to
be standard (Frail et al. 2001; Panaitescu \& Kumar 2001; Bloom,
Frail \& Kulkarni 2003; Berger, Kulkarni \& Frail 2003a). There are
two straightforward interpretations of this fact. One is that
different GRBs collimate the same total energy into different angular
openings (ranging from 1 to 30 degrees) (Rhoads 1999; Frail et
al. 2001). Another is that all GRBs have a universal jet shape with a
varying energy per solid angle which is of the form $\epsilon(\theta)
\propto \theta^{-2}$, where $\theta$ is the polar angle respect to
the jet axis (Rossi, Lazzati \& Rees 2002; Zhang \& \Mesz~2002a).
In principle, an individual GRB could also have other jet structure,
e.g. a Gaussian or even arbitrary function
(Zhang \& \Mesz~2002a). When the jet parameters are allowed to have
some dispersion around the mean values (which is demanded by the
data), one has a ``quasi-universal'' jet structure, and a power-law
jet structure is no longer a pre-requisite (Lloyd-Ronning, Dai \&
Zhang 2003). 

Recently the so-called X-ray flashes (XRFs, Heise et al.
2003; Kippen et al. 2003) are identified and proposed as a natural
extension of GRBs into a softer and fainter 
regime. Another intriguing empirical fact is that the GRB-XRF spectral
break energy ($E_p$) in the cosmic rest frame appears to be correlated
with the ``isotropic-equivalent'' energy (or luminosity) of the explosion
(Amati et al. 2002; Sakamoto et al. 2003; Lamb, Donaghy \& Graziani 2003;
Lloyd et al. 2000) according to
\be
E_p \sim 100~{\rm keV} \left(\frac{E_{iso}}{10^{52} ~{\rm
erg}}\right)^{1/2}.
\label{Ep}
\ee
Although the data sample in the XRF regime is currently very small
(because the faintness of XRFs hinders the detection of their possible
afterglows and the measurements of their redshifts), this intriguing
result, if confirmed by further future data, strongly suggests that
GRBs and XRFs are related events. A successful model should
therefore be able to interpret both GRBs and XRFs within a unified
framework. It is worth noticing that GRB 980425 has a very low
$E_{iso}$ ($\sim \sim 10^{48}$ ergs) but a relative large $E_p$ ($\sim
70$ keV), which is apparently inconsistent with eq.(\ref{Ep}). Because
it is atypical in other aspects as well, we do not interpret it within
the current framework. Since there is no unified terminology, in this
paper we call the combined population of GRBs and XRFs as GRB-XRF.

In this Letter, we do not discuss the nature of the $E_p \propto
(E_{iso})^{1/2}$ empirical law\footnote{We note however that this
information potentially contains important information to
constrain the nature of the GRB fireball (whether or not it is
magnetic-energy-dominated) and the location of the GRB prompt
emission (whether internal or external). See Zhang \&
\Mesz~(2002b) for more detailed discussions.}. Rather we
conjecture that it is a universal law for the majority of
GRB-XRFs, and evaluate its implications for various GRB-XRF jet
models. We will show that a quasi-universal Gaussian-like structured
jet model is compatible with the current prompt emission and afterglow
data of GRBs and XRFs. This is consistent with earlier suggestions 
that GRBs would be seen as X-ray transients when viewed at 
large angles (e.g. \Mesz~\& Rees 1997; Woosley, Eastman \& Schmidt 
1999 and subsequent work).

\section{Constraints on other jet models}

A direct implication of the $E_p \propto (E_{iso})^{1/2}$ law is
that it poses important constraints both on the specific
``universal'' structured jet model that assumes a power-law
structure with index 2 (or $k=2$ power-law model, Rossi et al. 2002;
Zhang \& \Mesz~2002a), and on the conventional ``uniform'' jet model
(Rhoads 1999; Frail et al.  2001)\footnote{The discussions here
are relevant when the prompt emission energy is roughly proportional 
to the kinetic energy, which is verified at least for GRBs since both 
geometry-corrected energy components are roughly standard.}. 

The constraint on the $k=2$ power-law universal model was raised by
Lamb et al. (2003). The argument is as follows: In the $k=2$ model,
we have $E_{iso} \propto \theta_v^{-2}$, where $\theta_v$ is the
viewing angle (since $\epsilon(\theta)$ is equivalent to $E_{iso}$
in the structured jet model). This implies $E_p \propto \theta_v^{-1}$.
Since the $E_p$ values range from $\simg 300$ keV in GRBs all the way
down to $\siml 3$ keV in XRFs, the viewing angles of XRFs need to be
two orders of magnitude larger than those of the GRBs. Even if XRFs
originate in standard jets viewed at the equator ($\theta_v \sim
90^{\rm o}$), all GRBs should have viewing angles less than $1^{\rm
o}$. The probability for a viewing angle $\theta_v$ is
$P(\theta_v) d\theta_v \propto \sin\theta_v d\theta_v$, thus a $k=2$
model greatly over-predicts the number of XRFs. This is in sharp
contrast with the observations, which indicate that XRFs represent
approximately 1/3 of the total GRB-XRF population (Lamb et
al. 2003).

The above argument may be regarded as supporting a uniform jet model
(Lamb et al., 2003), where the radiation is seen ``on-beam". However,
it does not address the afterglow data such as light-curve breaks,
and leads to energetic and progenitor number inconsistencies.
This is because in this model, even if XRFs correspond to
isotropic events, GRBs have to still be jets with typical opening
angles less than $1^{\rm o}$. In the standard afterglow model,
the bulk Lorentz factor evolves as $\Gamma(t) \simeq 6
(E_{52}/n)^{1/8} (t/\mbox{1 day})^{-3/8}(1+z)^{3/8}$, where $E_{52}$
is the isotropic kinetic energy of the fireball, $n$ is the
interstellar medium density, $t$ is the observer's time, and $z$ is
the redshift. Taking the standard view that the jet break time $t_j$
(around days) corresponds the epoch of $1/\Gamma(t) = \theta_j$
(Rhoads 1999), we get the constraint
\be
\frac{E_{52}}{n} \simeq 8.5\times 10^6 \left(\frac{t_j}{\mbox{1 day}}
\right)^3 \left(\frac{\theta_j}{1^{\rm o}} \right)^{-8}
\left(\frac{1+z}{2}\right)^{-3}~,
\ee
which requires an extremely large kinetic energy, or an extremely
low medium density, or both. This is in sharp contrast with current
afterglow analyses (Freedman \& Waxman 2001; Panaitescu \& Kumar 2001;
Berger et al. 2003a). The price of accepting this picture would be to
abandon the current afterglow theory completely, which has been well
tested and proven adequate to deal with the bulk of the afterglow data.
If, on the other hand, we take $E_{52} \sim 1$ as inferred from 
data (e.g. Berger et al. 2003a), this narrow-beam picture would imply
that GRBs have a total energy of order $10^{49}$ ergs. It would also
imply that the number of GRBs is similar to the number of SN Ib/c,
whereas a sample of SN Ib/c reveals only $3\%$ have radio afterglows
(Berger et al. 2003c). 

A variant of the uniform jet model is to interpret GRBs as on-beam
detections and XRFs as off-beam detections (Yamazaki, Ioka \&
Nakamura 2003a). This model may have the prospect of both interpreting
the correct XRF-to-GRB ratio and preserving the standard afterglow
model, but still requires GRB jets to have a large dispersion of
opening angles. Furthermore, the XRFs' afterglows in such a model
should resemble those in the ``orphan afterglows'', i.e., initially
rising and peaking at a time $t_{pk}$ when $1/\Gamma=\theta_v$, where
$\Gamma$ is the Lorentz factor within the uniform jet cone, and
$\theta_v$ is the viewing angle (e.g. Granot 
et al. 2002). Since $\theta_v > \theta_j$ in this model, one should
expect that $t_{pk}$ is typically larger than the typical $t_j$ in GRB
afterglows. The recent afterglow observations for XRF 030723 indeed
show an initial rising lightcurve (Huang et al 2003, and references
therein), but $t_{pk}$ is around 0.1 day, much smaller than the
typical $t_j$ for GRBs, which is typically several days. In order to
interpret the XRF as an off-beam GRB, the jet opening angle has to be
anomalously small. The lightcurve, however, is consistent with the
standard on-beam afterglow with the peak being due to crossing of the
typical synchrotron frequency across the optical band (e.g. Kobayashi
\& Zhang 2003). Similarly, the optical afterglow data for XRF 020903
(Soderberg et al. 2003) indicate that the lightcurve already decays
starting from 0.9 days, which is not consistent with an orphan
afterglow lightcurve.

\section{A quasi-universal Gaussian-like structured jet model}

Below we will argue that the current GRB-XRF prompt emission and
afterglow data are compatible with a model in which the GRB-XRF jets
have a quasi-universal structure, where the jet energy distribution
is axially symmetric and Gaussian-like, i.e., a jet which still has
a typical angle, with a mild variation of energy within this jet
typical angle and a rapid (e.g. exponential) decrease of energy outside
the typical angle (Zhang \& \Mesz~2002a; Lloyd-Ronning et al. 2003).
For analytical purposes, we approximate the angular distribution of
the jet energy as
\be
\epsilon(\theta) = \epsilon_0 e^{-\frac{\theta^2}{2\theta_0^2}}.
\label{Gaussian}
\ee
The initial Lorentz factor should also have an angular dependence,
and we take it as a free function, since it is in principle independent
and since it does not directly influence the GRB-XRF population
study\footnote{It, however, influences the rising time of the afterglow
lightcurve. See below for further discussion.}. The motivation for
introducing a Gaussian-like jet are dual. First, it preserves a
characteristic angle for the jet, $\theta_0$, which is more
consistent with the jet structure generated in the numerical
simulations for the collapsar model (Zhang, Woosley \& MacFadyen
2003a; Zhang, Woosley \& Heger 2003b). It also alleviates the
divergence problem of a power-law model at small angles by introducing
a smooth functional profile within the characteristic jet angle. 
Second, at larger angles (beyond $\theta_0$), the energy decrease
is steeper than the power law model, so that in order to get
an XRF whose $E_{iso}$ is $(10^2-10^4)$ times lower than the typical
GRB $E_{iso}$, one only needs to have a viewing angle $\theta_v \sim
(3-4)\theta_0$. This greatly reduces the predicted number of XRFs.

We perform a simple Monte Carlo simulation to verify this ansatz. Since
it is unreasonable to expect that all GRB-XRFs are exactly the same,
we allow some scatter of the model parameters. Such scatter is also
needed (Lloyd-Ronning et al. 2003) to reproduce the observed
$E_{iso}-\theta_j$ correlations from the afterglow data (Frail et
al. 2001; Bloom et al. 2003), even for the $k=2$ power-law jet
model. We randomly generate 10000 bursts, each
of which has a jet structure in the form of eq.(\ref{Gaussian}). We
define the total energy within the jet as $E_j =2\pi \int_0^{\pi/2}
\epsilon(\theta) \sin\theta d\theta$. For small angles, this is $2\pi
\epsilon_0 \theta_o^2$ (Zhang \& \Mesz~2002a), but here we use a more
rigorous integration which is also applicable for large angles. We
assume that the distributions of both $E_j$ and $\theta_0$ are
lognormal. For each realization, we use the generated $E_j$ and
$\theta_0$ to derive $\epsilon_0$ and to obtain the jet structure
according to (eq.[\ref{Gaussian}]). For each burst, we also generate a
viewing angle with the probability $P(\theta_v) d\theta_v \propto
\sin(\theta_v) d\theta_v$. The corresponding $4\pi\epsilon(\theta_v)$ is
then assigned to be the $E_{iso}$ for that particular burst. The cosmic
rest frame $E_p$ of that burst is generated via eq.(\ref{Ep}) with a
lognormal variation (with a standard deviation of 0.3), and the
observed peak energy is $E_p(obs)=E_p/(1+z)$. The redshift
distribution of the bursts is assumed to trace the cosmic star-forming
rate (Rowan-Robinson 1999). Standard cosmological parameters are
adopted, i.e., $H_0=70~{\rm km~s^{-1}~Mpc^{-1}}$, $\Omega_m=0.3$ and
$\Omega_\Lambda=0.7$. No redshift-evolution for the burst parameters
are assumed.
We then assign a redshift for each burst, and calculate the distance
and the energy fluence for that burst. Finally we place a fluence
truncation of $5\times 10^{-8}~{\rm erg~cm^{-2}}$ to reflect the
instrument sensitivity limit. This number matches the faintest HETE-2
XRF (Fig.1 of Lamb et al. 2003), and we regard it as reflecting HETE-2
sensitivity. (We have experimented with changing the fluence
truncation limit. In general, the total number of the detectable
GRB-XRFs increases with a deeper fluence truncation, and the fraction
of XRFs from the whole population increases mildly.) The simulated
bursts are 
plotted in the $E_{p}(obs)$-fluence plane (Fig.1), together with
the BeppoSAX and HETE-2 prompt emission data.

The model can be also compared to the afterglow light curve break data
(Bloom et al. 2003). In a structured jet, if the jet structure is
steep enough (e.g. for the power law jets), the viewing angle defines
the jet break time (Rossi et al. 2002; Zhang \& \Mesz~2002a; Kumar \&
Granot 2003; Granot \& Kumar 2003; Panaitescu \& Kumar 2003; Wei \&
Jin 2003; Salmonson 2003). For a Gaussian jet, although the viewing
angle defines the jet break time for $\theta_v > \theta_0$, the jet
structure is only mild within the typical angle $\theta_0$, so that
for $\theta_v < \theta_0$, it is $\theta_0$ rather than $\theta_v$
which defines the jet break time (Kumar \& Granot 2003). In our
simulation, we assign a jet break angle $\theta_j=\mbox{max}(\theta_0,
\theta_v)$ for each burst\footnote{The result does not differ too much
from the present simulation if we assumed that $\theta_v$ is exclusively
the agent which defines the jet break (e.g. Lloyd-Ronning et al. 2003).
The reason is that the probability for small viewing angles (which
make the difference) is small.}.
The fluence-truncated bursts are plotted in the $E_{iso}-\theta_j$ plane
(Fig.2), with the afterglow data.

Figures 1 and 2 show our  simulation results. The parameters
adopted are $\log (E_j/\mbox{1 erg})= 51.1\pm 0.3$ and $\log
(\theta_0/\mbox{1 rad}) = -1.0 \pm 0.2$, where the error is the
standard deviation of the lognormal distribution. This corresponds
to a quasi-standard explosion energy of
$E_j= 1.3^{+1.2}_{-0.6} \times 10^{51}$ ergs, and a quasi-standard
jet configuration with the typical opening angle of
$\theta_0=5.7^{+3.4}_{-2.1}$ degrees. This is in excellent agreement
with collapsar simulations (Zhang et al. 2003a). Figure 1 shows
that this model is compatible with the prompt emission data
collected by BeppoSAX and HETE-2. Defining XRFs as those events
with $E_p(obs) < 25$ keV (Soderberg et al. 2003), the simulated number
of XRFs is roughly 1/3 of the whole GRB-XRF population within the
fluence-truncated observed sample, as in the observations.
Figure 2 shows that the same set of parameters is compatible
with the jet angle data within the framework of the standard
afterglow model. 

Among 10000 simulated bursts, 702 GRB-XRFs survive the
fluence-selection effect. Since this takes viewing angle effects into
account, the simulation suggests that the true-to-observed number ratio
for the entire combined GRB-XRF population is about 14 (or 21 for GRB
plus X-ray-rich bursts, or 42 for GRBs alone). The ratio should decrease
for higher sensitivities. This number is much smaller than the
beaming correction factor ($\sim 500$) in the uniform jet model for
the GRB population (Frail et al. 2001; van Putten \& Regimbau
2001), suggesting that the number of GRB-XRF progenitors required is
smaller than previously thought.

Kumar \& Granot (2003) have performed a numerical
hydrodynamical modeling of the evolution of a Gaussian-jet, which
reveals a substantial energy redistribution in the jet structure during
its evolution. The resultant lightcurves (their Fig.5a) are consistent
with most of the jet break data. For $\theta_v < \theta_0$, the
lightcurves are similar to those of a uniform jet model which is
consistent with the data. For $\theta_v > \theta_0$, a jet break is
visible around the time of $\Gamma(\theta_v) \sim \theta_v^{-1}$. In
their calculations, the structure of $\Gamma(\theta_v)$ is also taken
as a Gaussian, so that when $\theta_v \gg \theta_0$, the lightcurves
become analogous to those of orphan afterglows characterized by a
late rising. However, if the initial Lorentz factor at large viewing
angles is high, i.e., $\Gamma_0(\theta_v) \gg \theta_v^{-1}$, the
rising segment of the 
lightcurve would be shifted to earlier times, followed by a
decaying lightcurve (Zhang \& \Mesz~2002a). For $\theta_v \gg
\theta_0$, a lightcurve bump may be expected for $\theta_v \gg
\theta_0$ when the jet axis becomes visible (J. Granot, 2003, personal
communication), since the main power at the jet axis is initially not
in the relativistic light cone. 
However, the energy redistribution effect (e.g. Fig.1 of Kumar \& Granot
2003) may smear the bump to be less significant. In any case, XRFs in
the current model only imply $\theta_v$ to be at most $4 \theta_0$.
Observationally, light curve  breaks are currently well monitored mainly
for those bursts whose break time is typically days, which corresponds
to small jet angles (i.e. $\theta_v \siml \theta_0$). The afterglow data
for the recent XRF 030723 (Huang et al. 2003 and references therein), on
the other hand, shows a lightcurve re-brightening after 10 days.
For an initial Gaussian jet (in the prompt phase), the jet
structure would evolve during the afterglow phase because of a
pole-to-equator energy flow. For a large viewing angle corresponding
to an XRF, the late-time afterglow energy in the viewing direction
could be much larger than the energy in the prompt phase.
This is consistent with the recent radio afterglow observation for XRF
020903 (Soderberg et al. 2003). We therefore
conclude that the model discussed here is not inconsistent with all
the current GRB-XRF afterglow data.

\section{Discussion}

We have argued that in order to incorporate both the prompt emission
and afterglow data for GRBs and XRFs within a unified theoretical
framework, both the simple $k=2$ power-law universal jet model and the
on-beam uniform jet model encounter difficulties. With a Monte Carlo
simulation, we show that the current data are compatible with a
quasi-universal Gaussian-like structured jet model.

Other possible models may still interpret the data. First, a
uniform jet also produces an equivalent isotropic energy outside the
jet cone due to the off-beam Doppler effect. So an off-beam model for
XRFs (Yamazaki et al. 2003a) does not over-generate XRFs. However, the
decline of the isotropic energy at larger viewing angles is even
steeper than exponential (e.g. Yamazaki, Yonetoku \& Nakamura
2003b). So one can only generate XRFs at viewing angles slightly
larger than the jet opening angle. This tends to {\em under-generate}
XRFs. The afterglow issue as discussed in \S2 also poses some
constraints on this model. In any case, a population simulation like
the one presented here is needed to validate that model. Second, it is
possible 
that some or even all GRB-XRF jets include two (or more) components
(Li \& Chevalier 1999; Frail et al. 2000; Lipunov, Postnov \&
Prokhorov 2001; Berger et al. 2003b; Sheth et al. 2003;
Huang et al. 2003), as reproduced from the collapsar models (Zhang et al.
2003a, 2003b; \Mesz \& Rees 2001; Ramirez-Ruiz, Celotti \& Rees
2003). In such a model, the jet structure
may be approximated as a superposition of two (or more) uniform or
Gaussian components. Such a model may be also able to reproduce the
GRB-XRF data. However, the afterglow lightcurves could show distinct
signatures under certain conditions (Berger et al. 2003b; Huang et al.
2003). The lack of such signatures in most of the afterglow lightcurves
poses constraints on the parameters for such two-component structured
jet models. Third, we do not exclude the possibility that a $k=2$ power
law structure extends only up to a certain angle, above which the jet
energy has a much steeper decline (Zhang \&
\Mesz~2002a). However, the current Gaussian model is the simplest among
these, and can accommodate the widest range of prompt emission and
afterglow data for the majority of GRBs and XRFs. More detailed Monte
Carlo simulations for this as well as other models and their
comparison to a wider set of detailed prompt emission and afterglow
data will be presented elsewhere (Dai \& Zhang 2003, in preparation,
see also Liang, Wu \& Dai 2003).

\acknowledgements
We are grateful to J. S. Bloom, D. A. Frail, J. Granot, S. Kobayashi,
A. Panaitescu, E. Ramirez-Ruiz, G. R. Ricker and S. E. Woosley for
discussions/comments, and to partial support through NASA NAG5-13286,
NAS8-01128, and the Monell Foundation.

\begin{figure}[ht]
\plotone{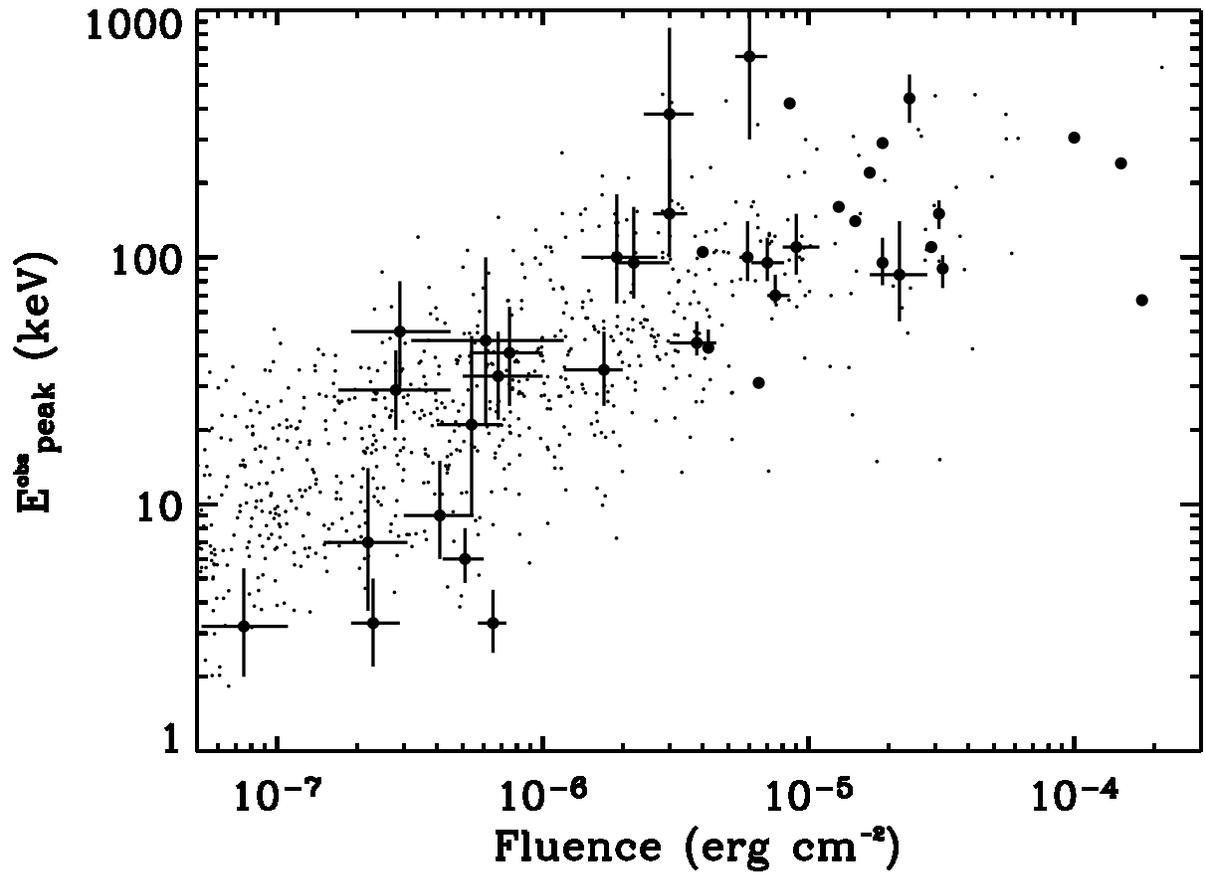}
 \caption{ The $E_p(obs)$-fluence diagram of the simulated GRB-XRFs
(small dots) as compared with the HETE-2 and BeppoSAX data (heavy
dots with error bars, data from Lamb et al. 2003).
 \label{fig1}}
 \end{figure}

\begin{figure}[ht]
\plotone{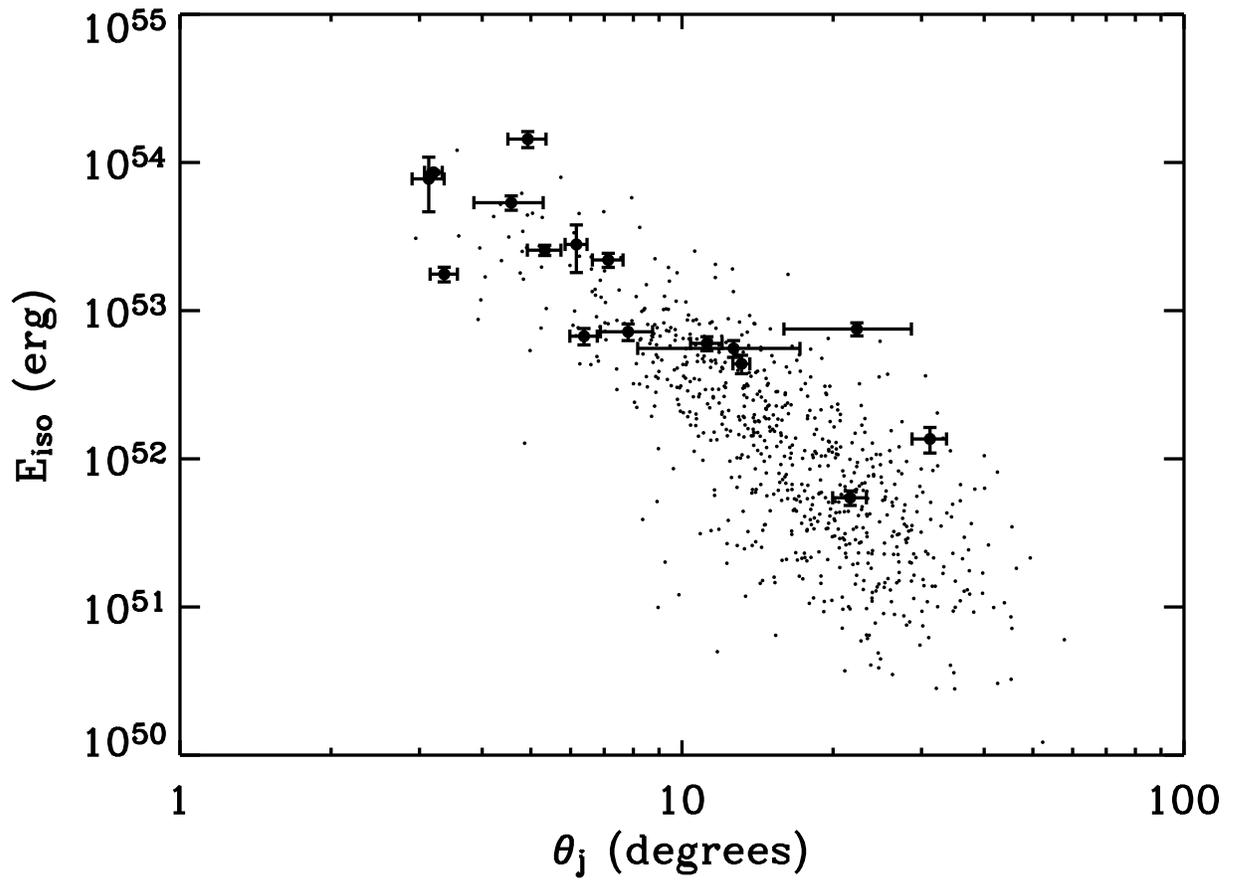}
 \caption{ The $E_{iso}-\theta_j$ diagram of the simulated GRB-XRFs
(small dots) as compared with the afterglow jet break data (heavy dots
with error bars, data from Bloom et al. 2003).
 \label{fig2}}
 \end{figure}

\end{document}